\documentclass{80SA}
\usepackage{bm,times}
\usepackage{amsmath}
\usepackage{latexsym}
\title{Dynamical Mean-Field Study of Metamagnetism in Heavy Fermion Systems}
\author{Takemi {\sc Yamada}\thanks{E-mail address: takemi@phys.sc.niigata-u.ac.jp} and Yoshiaki {\sc \=Ono}}
\inst{ Department of Physics, Niigata University, Ikarashi, Nishi-ku, Niigata, 950-2181, Japan }

\abst{
We investigate the metamagnetism in the periodic Anderson model with the $\bm{k}$-dependent $c-f$ mixing by using the dynamical mean-field theory combined with the exact diagonalization method. It is found that both effects of the $\bm{k}$-dependent $c-f$ mixing and strong correlation due to the Coulomb interaction between $f$ electrons are significant for determining both the magnetization and the mass enhancement factor. For the case away from the half-filling, the results is consistent with the metamagnetic behavior observed in CeRu$_2$Si$_2$. 
}

\kword{heavy fermion system, dynamical mean-field theory, periodic Anderson model, metamagnetism, CeRu$_2$Si$_2$}

\begin{document}
\maketitle

\section{Introduction}
In the heavy fermion systems, the conduction $(c)$ electrons mix with the almost localized $f$ electrons, and form the strongly renormalized quasiparticles, which have effective masses of 100$\sim$1000 times larger than the bare electron masses. Because of this strong renormalization, the heavy fermion quasiparticles often exhibit sensitive behavior to the external perturbation such as magnetic field $H$ or pressure $P$. For example, the metamagnetism is one of the most remarkable features of the heavy fermion systems. The heavy fermion compound CeRu$_2$Si$_2$ exhibits an abrupt increase in the magnetization at an applied magnetic field $H_{m}\sim 7.7~{\rm T}$\cite{exp1} where the other physical quantities also show various anomalies without any symmetry breaking or magnetic phase transition. This is known as a ``heavy fermion metamagnetism'' and has been studied extensively in past decades.\cite{exp1,exp2,exp3,exp4,exp5}

To understand the physics of the heavy fermion systems including the metamagnetism from a microscopic viewpoint, we have to consider the effects of the $c-f$ mixing and the strong correlation due to the Coulomb interaction between $f$ electrons which are crucial for the heavy fermion properties. The Kondo lattice model (KLM) and periodic Anderson model (PAM) are fundamental models for studying the heavy fermion properties and many theoretical studies have been carried out on the basis of these models.\cite{Gut,SB,NCA,1/N} Since it is difficult to solve these models exactly due to the many-body problem, some approximations have been required to solve these models, such as the Gutzwiller approximation\cite{Gut}, slave boson mean-field\cite{SB}, non-crossing approximation\cite{NCA} and $1/N$ expansion\cite{1/N}.

Dynamical mean-field theory (DMFT) has been known as one of the most powerful method to describe the strongly correlated electron systems, since the DMFT includes the local quantum fluctuation effects which plays an essential role for the heavy fermion systems.\cite{DMFT1,DMFT2,DMFT3,DMFT4,DMFT5,DMFT6} The early DMFT studies on the PAM for the heavy fermion systems have succeeded in describing the strongly renormalized quasiparticles and the magnetization process.\cite{DMFT3,DMFT4,DMFT5} These studies were restricted to the case with the half-filling and with a $\bm{k}$-independent $c-f$ mixing, where the Kondo insulator state is realized. As for the metallic case, the PAM away from the half-filling has been studied by using the DMFT in combination with the iterated perturbation theory (IPT).\cite{DMFT6} In contrast to the experiment of CeRu$_2$Si$_2$, however, the metamagnetism occurs together with the ferromagnetic transition in ref. 15). Hence, we need more realistic model for the study on the metamagnetism of CeRu$_2$Si$_2$.

In addition to the strong correlation effect, the $\bm{k}$-dependence of $c-f$ mixing is considered to be important for the metamagnetism of CeRu$_2$Si$_2$\cite{Gut,SB,1/N}, where the $\bm{k}$-dependence originates from the local $f$ state in the crystal electric field (CEF)\cite{cf-mix} and yields the specific density of states (DOS) with the three peak structure. In this paper, we investigate the heavy fermion metamagnetism on the basis of the PAM with the $\bm{k}$-dependent $c-f$ mixing by using the DMFT combined with the ED (DMFT+ED) method. 

\section{Model and formulation}
Our model Hamiltonian is the PAM with $\bm{k}$-dependent $c-f$ mixing, which consists of the $c$ electron term $H_{c}$, $f$ electron term $H_{f}$ and the $c-f$ mixing term $H_{cf}$ given by
\begin{align}
H&=H_{c}+H_{f}+H_{cf}\label{eq:H}\\
&H_{c}=\sum_{\bm{k}\sigma}\epsilon_{\bm{k}}n_{\bm{k}\sigma}^{c},\label{eq:Hc}\\
&H_{f}=\sum_{im}\epsilon_{fm}n^{f}_{im}+U\sum_{i}n^{f}_{i+}n^{f}_{i-},\label{eq:Hf}\\
&H_{cf}=\sum_{\bm{k}m\sigma}\left(V_{\bm{k}m\sigma} f_{\bm{k}m}^{\dagger}c_{\bm{k}\sigma}+h.c.\right),\label{eq:Hcf}
\end{align}
where $c_{\bm{k}\sigma}^{\dagger}$ is a creation operator for a $c$ electron with wave vector $\bm{k}$ and spin $\sigma=\uparrow,\downarrow$ and $f_{im}^{\dagger}$ is that for a $f$ electron at site $i$ with the Kramers doublet $m=\pm$ in the CEF groundstate. $U$ is the Coulomb interaction between $f$ electrons and $V_{\bm{k}m\sigma}$ is the $c-f$ mixing. $~\epsilon_{\bm{k}}$ and $\epsilon_{fm}$ are energies for the $c$ and the $f$ electrons, respectively. For simplicity, the $c$ electron is assumed to be a free electron with a rectangular DOS and the band width is $2D$. The CEF groundstate in the case of CeRu$_2$Si$_2$ is approximately given by $J_z=\pm 5/2$ Kramers doublet where the $c-f$ mixing is given by $V^{2}({\bm{k}})=\sum_{m}|V_{\bm{k}m\sigma}|^{2}=\sum_{\sigma}|V_{\bm{k}m\sigma}|^{2}=\frac{15}{8}V_{cf}^{2}(1-{\rm cos}^{2}\theta_{\bm{k}})^2$ with ${\rm cos}\theta_{\bm{k}}=k_z/|\bm{k}|$.\cite{cf-mix} Here, we also assume that the magnetic field $H$ is included only on the $f$ electron $\epsilon_{fm}=\epsilon_{f}-m H$ with $m=\pm$ because the $g$ value for $f$ electrons is much larger than that for $c$ electrons.

We have solved the above Hamiltonian using the DMFT+ED method. In this method, the lattice Hamiltonian is mapped onto an effective impurity Anderson model, which includes the so-called Weiss field parameters (WFPs) and is solved with the Lanczos ED algorithm. The WFPs are determined so as to satisfy the self-consistent equation given by,
\begin{equation}
{G}_{m}(i\epsilon_{\nu})=\frac{1}{N}\sum_{\bm{k}}\left(i\epsilon_{\nu}-\epsilon_{fm}-\Sigma_{m}(i\epsilon_{\nu})-\frac{V^{2}({\bm{k}})}{i\epsilon_{\nu}-\epsilon_{{\bm k}}}\right)^{-1},
\end{equation}
where ${G}_{m}(i\epsilon_{\nu})$ and $\Sigma_{m}(i\epsilon_{\nu})$ are the Green's function and self-energy of the effective impurity Anderson model at $T=0$, respectively. In eq. (5), $i\epsilon_{\nu}=i(2\nu+1)\pi\bar{\beta}~(\nu=0,1,\cdots,n_{max})$, and $\bar{\beta}$ is not an inverse temperature but a parameter which determines the energy resolution in the DMFT.\cite{DMFT2}

In this study, we set the parameters as follows : the half of the $c$ band width $D=2$, $c-f$ mixing strength $V_{cf}=0.5$, the frequency cutoff $n_{max}=4000$ and $\bar{\beta}=600-800$. We use the effective impurity Anderson model with the effective bath sites $N_s=8$ and confirm that the results are almost unchanged for $N_s=6-10$. In the present paper, we will show the results for the half-filling case with  $\langle n\rangle=\langle n_f\rangle+\langle n_c\rangle=2.0$ and the away from half-filling case with $\langle n\rangle=1.7$. 

\section{Results}
Figure 1 shows the renormalized DOS $\rho_f(\epsilon)=-\frac{1}{\pi}{\rm Im}G_{m}(\epsilon+i0_{+})$ near the fermi level $\varepsilon=0$ for several values of $U$ at $\langle n\rangle=2.0$. In this calculation, we perform the analytic continuation of $\Sigma_{m}$ from the imaginary frequency to the real frequency by using the Pad$\acute{\rm e}$ approximation. When $U=0$, we can see the three peak structure of the DOS which is due to the $\bm{k}$-dependent $c-f$ mixing, where we refer the peaks in the left, center and right as peak1, peak2 and peak3 as shown by arrows in Fig. 1, respectively. We note that there is no hybridization gap in the present model where the $\bm{k}$-dependent $c-f$ mixing vanishes around ${\rm cos}\theta_{\bm{k}}\sim 1$. When $U$ increases, the renormalized quasiparticle band width decreases with decreasing the integrated spectral weight of the quasiparticle band, and then the peak1 and peak3 shift to the fermi level as shown in Fig. 1. In addition, we also observe broad peaks correspond to the Hubbard like bands around $\epsilon_{f}=-U/2$ and $\epsilon_{f}+U=U/2$ (not shown).

To see the $U$-dependence of the correlation effect more explicitly, we plot the renormalization factor defined by $Z_{m}=\left(1-\frac{d}{d\epsilon}{\rm Re}\Sigma_{m}(\epsilon)\Bigr|_{\epsilon=0}\right)^{-1}$ for $\langle n\rangle=2.0$ and $\langle n\rangle=1.7$ in Fig. 2, where $Z_{m}^{-1}$ is nothing but the mass enhancement factor: $m^{*}/m=Z_{m}^{-1}$. In the cases with both of $\langle n\rangle=2.0$ and $\langle n\rangle=1.7$, $Z_{m}$ decreases with increasing $U$ and the heavy fermion state is realized for $U\hspace{0.3em}\raisebox{0.4ex}{$>$}\hspace{-0.75em}\raisebox{-.7ex}{$\sim$}\hspace{0.3em} 2.0$, where $Z_{m}^{-1}$ becomes large as 10 or more. Here, we note that, for $\langle n\rangle=2.0$, we consider the particle-hole symmetry case with $\epsilon_{f}=-U/2$, while, for $\langle n\rangle=1.7$, $\epsilon_{f}$ is fixed at $\epsilon_{f}=-1.7$ measured relative to the $c$ band center.

\begin{figure}[t]%-----------------------------------------------------
\begin{center}
\includegraphics[width=6.25cm]{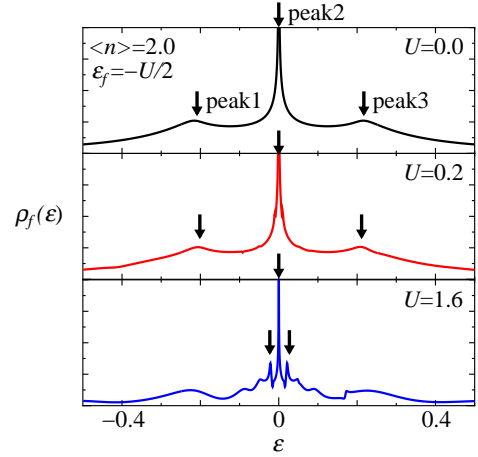}
\vspace{-0.5cm}
\caption{The renormalized DOS near the fermi level $\varepsilon=0$ for several values of $U$ at $\langle n\rangle=2.0$ and $\epsilon_{f}=-U/2$.}
\end{center}
\end{figure}%-----------------------------------------------------------
\begin{figure}[t]%-----------------------------------------------------
\vspace{-0.75cm}
\begin{center}
\includegraphics[width=6.00cm]{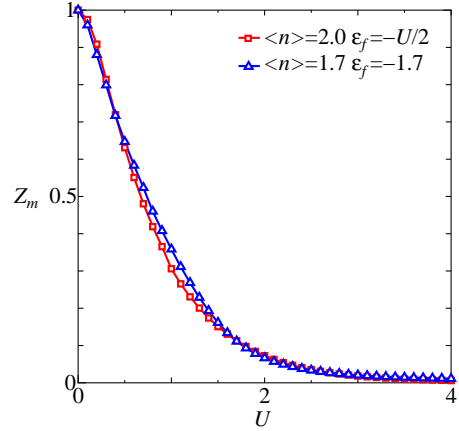}
\vspace{-0.5cm}
\caption{The $U$ dependence of the renormalization factor $Z_{m}$ for $\langle n\rangle=2.0$ with $\epsilon_{f}=-U/2$ ($\Box$) and $\langle n\rangle=1.7$ with $\epsilon_{f}=-1.7$ ($\triangle$).}
\end{center}
\end{figure}%-----------------------------------------------------------

Fig. 3 (a) shows the magnetization curves $M=\langle n^{+}_{f}\rangle-\langle n^{-}_{f}\rangle$ for several values of $U$ at $\langle n\rangle=2.0$. We can see nonlinear magnetization curves attributed to the $\bm{k}$-dependent $c-f$ mixing. The metamagnetic behavior observed in CeRu$_2$Si$_2$ is not obtained in the half-filling case. From the magnetization curve, we also calculate the differential susceptibility $dM/dH$ (not shown) and find that $dM/dH$ shows a peak at a critical magnetic field where the peak1 and peak3 of the DOS cross the fermi level. When $U$ increases, the critical magnetic field decreases with decreasing the quasiparticle band width shown in Fig.1. We note that the shape of the renormalized DOS are largely depends on $H$ especially around the Hubbard like bands which dominantly contribute to $M$ for the case with large $U$ where the $f$ electrons are almost localized.

The $H$-dependence of the mass enhancement factor $Z_{m}^{-1}$ is plotted in Fig. 3 (b) for several values of $U$ at $\langle n\rangle=2.0$, where $Z_{+}=Z_{-}$ even for $H\ne 0$ because of the the particle-hole symmetry. When $H$ increases, $Z_{m}^{-1}$ decreases especially for the large $U$ case, where the heavy fermion state is largely suppressed due to $H$. 

Fig. 4 (a) shows the magnetization curves for several values of $U$ at $\langle n\rangle=1.7$. In this case, the fermi level with $H=0$ is located between the peak1 and the peak2. When $H$ increases, the differential susceptibility shows a peak at a critical magnetic field $H_{m}$ where the peak1 and the peak2 cross the fermi level almost at the same time. Then we observe the heavy fermion metamagnetic behavior at $H=H_{m}$. In the intermediate correlation regime with $U\sim 1$, the saturated magnetization above $H_m$ is almost a half of the full moment and such behavior is consistent with the result observed in CeRu$_2$Si$_2$\cite{exp1}. In the strong correlation regime with $U\sim 4$, the magnetization curve is similar to the case with the localized spin system where the metamagnetic behavior is suppressed.

We also show the $H$-dependence of the mass enhancement factor $Z_{m}^{-1}$ at $\langle n\rangle=1.7$ for $U=0.8$ and 2.4 in Fig. 4 (b). For both cases, $Z^{-1}_{+}$ (majority component) is larger than $Z^{-1}_{-}$ (minority one) as previously obtained for the PAM with $\bm{k}$-independent $c-f$ mixing\cite{DMFT6}. In addition, we also observe a shallow peak of $Z^{-1}_{+}$ at $H=H_{m}$ due to the cooperative effect of the strong correlation and the $\bm{k}$-dependent $c-f$ mixing. 

\begin{figure}[t]
\begin{center}
\includegraphics[width=6.30cm]{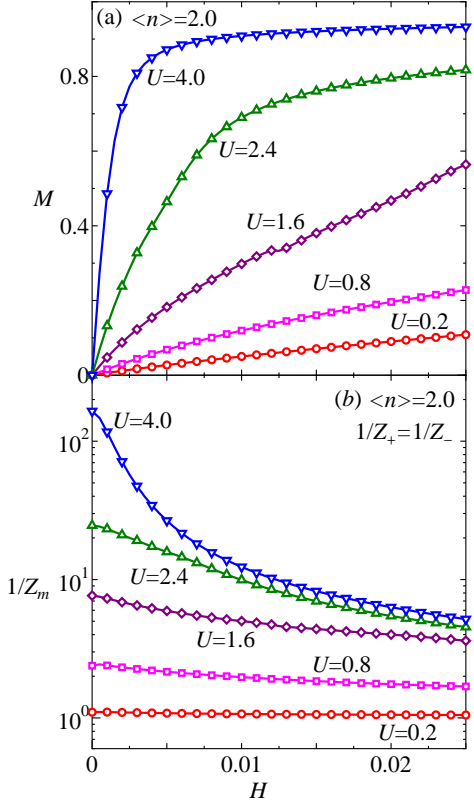}
\vspace{-0.5cm}
\caption{The magnetization (a) and mass enhancement factor $Z_{m}^{-1}$ (b) for several values of $U$ at $\langle n\rangle=2.0$ with $\epsilon_{f}=-U/2$.}
\end{center}
\end{figure}
\begin{figure}[t]
\begin{center}
\includegraphics[width=6.30cm]{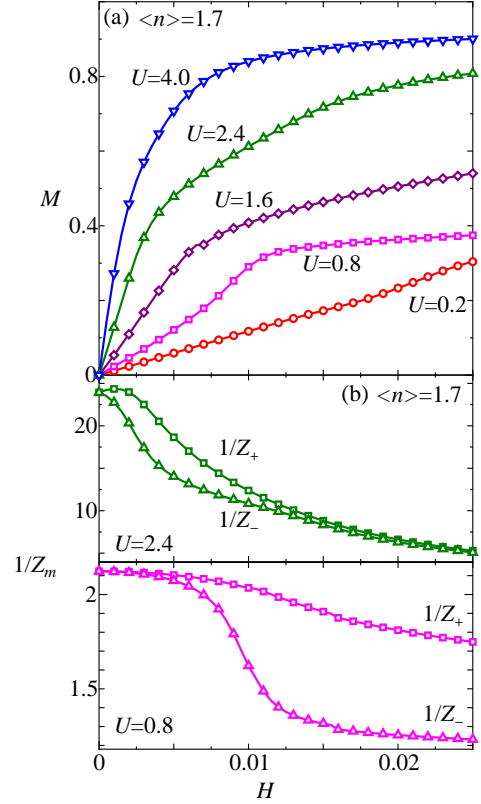}
\vspace{-0.5cm}
\caption{The magnetization $M$ (a) and mass enhancement factor $Z_{m}^{-1}$ (b) for several values of $U$ at $\langle n\rangle=1.7$ with $\epsilon_{f}=-1.7$.}
\end{center}
\end{figure}

\section{Summary and discussion}
In this paper, we have studied the heavy fermion metamagnetism on the basis of the PAM with the $\bm{k}$-dependent $c-f$ mixing under the external magnetic field $H$, and have found that the both effects of the $\bm{k}$-dependent $c-f$ mixing and the correlation due to the Coulomb interaction between $f$ electrons are crucial for the metamagnetism. For $\langle n\rangle=1.7$, the results with the intermediate correlation is consistent with the metamagnetic behavior observed in CeRu$_2$Si$_2$.

Experimentally, the heavy fermion metamagnetism is accompanied by a large magnetostriction\cite{volume}.  In the present study, we studied the case of the constant volume, i.e. the constant parameters $\epsilon_{\bm{k}},\epsilon_{f},V_{cf}$. When we vary the volume (parameters) under a constant pressure, we may obtain such magnetostriction together with an enhanced metamagnetic behavior.\cite{1/N} The explicit calculation with the constant pressure together with suitable parameter set of $\epsilon_{\bm{k}},\epsilon_{f},V_{cf}$ so as to reproduce the experimental observation in CeRu$_2$Si$_2$ will be shown in a subsequent paper. 

\section{Acknowledgment}
This work was partially supported by the Grant-in-Aid for Scientific Research from the Ministry of Education, Culture, Sports, Science and Technology of Japan. T. Y. is a research fellow of the Japan Society for the Promotion of Science.

\end{document}